A new type of photonic transistor concept based on a photo-ferroelectric crystal gated at the onset of a bandgap energy, is demonstrated. The photovoltaic charge generation impacts the internal electric field resulting in a coherent-less optical computing possibility. Thanks to ferroelecticity, the approach is also able to demonstrate an all-optical and re-writable memory effect.

# Photovoltaic-Ferroelectric Materials for the Realization of All-Optical Devices

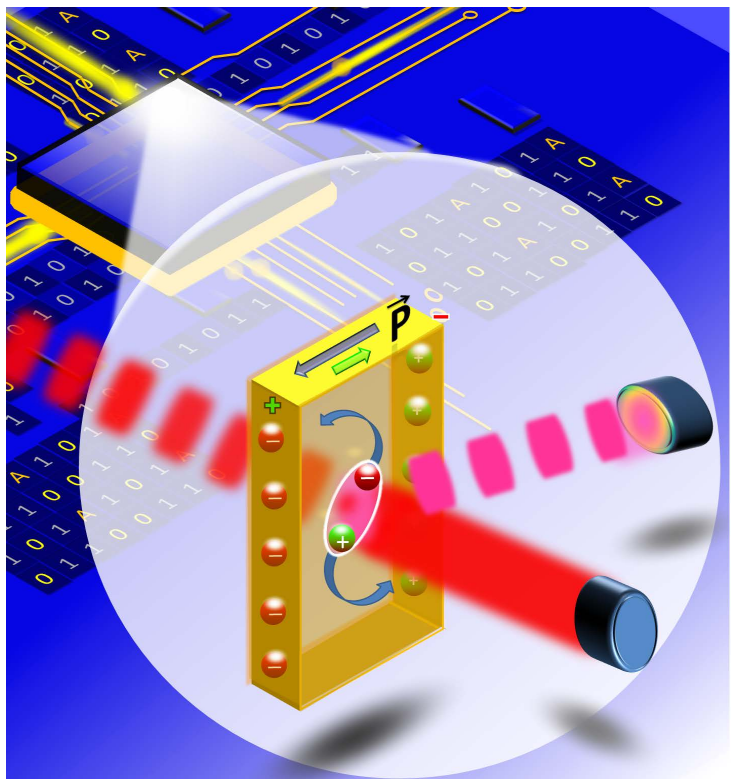

A. Makhort, R. Gumeniuk, J.-F. Dayen, P. Dunne, U. Burkhardt, M. Viret, B. Doudin, B. Kundys*

# Photovoltaic-Ferroelectric Materials for the Realization of All-Optical Devices

*A. Makhort, R. Gumeniuk, J.-F. Dayen, P. Dunne, U. Burkhardt, M. Viret, B. Doudin, B. Kundys\**

**Following how the electrical transistor revolutionized the field of electronics, the realization of an optical transistor in which the flow of light is controlled optically should open the long-sought era of optical computing and new data processing possibilities. However, such function requires photons to influence each other, an effect which is unnatural in free space. Here it is shown that a ferroelectric and photovoltaic crystal gated optically at the onset of its bandgap energy can act as an optical transistor. The light-induced charge generation and distribution processes alter the internal electric field and therefore impact the optical transmission with a memory effect and pronounced nonlinearity. The latter results in an optical computing possibility, which does not need to operate coherently. These findings advance efficient room temperature optical transistors, memristors, modulators and all-optical logic circuits.**

## 1. Introduction

The electrical transistor is the fundamental building block of modern electronics which uses voltage gate control over the flow of electrical charges. In particular, electric gain in current with respect to the ground state is a technologically desired property. In the optical counterpart of this function, it must be possible to increase a photon flow or modulate its transmission by purely optical means. Such a property under ambient conditions remains largely hypothetical, although finding efficient optical transistors could revolutionize conventional computers. Recent reports exploited various environments and circumstances focusing on quantum effects in atomic,[1–8] molecular,[9] quantum dots,[10–12] polaritons[13,14] or graphene[15]- based systems, including electromagnetically induced transparency effects.[16,17] To be technologically compatible the all-optical gating of the light transmission must operate at room temperature with low light power and demonstrate logic functionality in solid state devices. There are also important application criteria to meet such as input-output insulation, cascadability, logic-level restoration, etc.[18] Thus, new simple and inexpensive device paradigms demonstrating efficient optical gating under ambient conditions are highly required. Because the mechanism by which a material interacts with light depends on its microscopic building blocks, an optical transistor can be realized based on a system that changes its intrinsic environment under illumination. Here we demonstrate the realization of this hypothesis in a cavity-free approach that does not need to operate coherently. The demonstration is realized using a ferroelectric (FE) crystal, possessing also photovoltaic (PV) properties. The photoferroelectric compounds are remarkable materials having potential for an increased multi-functionality.[19–21] The photovoltaic properties in some FE compounds[22–25] offer an efficient interplay between light induced photo-carrier generation and non-zero intrinsic electric field in the sample. Such interplay can lead to nonlinear optical response required for changing the light propagation. Herein we report that a light transmission through a photovoltaic-ferroelectric crystal can be modulated optically via the amount of charges generated and distributed during the bulk photovoltaic effect (BPVE).[22–24,26]

A. Makhort, J.-F. Dayen, P. Dunne, B. Doudin, B. Kundys
Université de Strasbourg CNRS,
Institut de Physique et Chimie des Matériaux de Strasbourg
UMR 7504, 23 rue du Loess, Strasbourg F-67000, France E-
\*mail: kundys[at]ipcms.unistra.fr

R. Gumeniuk
Institut für Experimentelle Physik
TU Bergakademie Freiberg
Leipziger Str. 23, Freiberg 09596, Germany

U. Burkhardt
Max Planck Institut für Chemische Physik fester Stoffe
Nöthnitzer Str. 40, 01187 Dresden, Germany

M. Viret
SPEC
CEA
CNRS
Université Paris-Saclay
Gif-sur-Yvette 91191, France

## 2. Results and Discussions

### 2.1. Optical Gating

The appropriate crystal of $Pb[(Mg_{1/3}Nb_{2/3})_{0.70}Ti_{0.30}]O_3$ was selected from the well-know family of the PMN–$x$PT complex optimized for piezoelectricity.[27,28] These compounds exhibit a rich phase diagram[29] and exceptional optical properties reaching ~70% transparency across a broad spectral range spanning from 0.4 to 6 microns.[30,31] Depending on the exact phase diagram region, some of these crystals were reported to be photovoltaic[32,33] or become such after doping.[34] Figure 1a illustrates the schematics of the experiment where a $Pb[(Mg_{1/3}Nb_{2/3})_{0.70}Ti_{0.30}]O_3$ single crystal with a structure at the border of the morphotropic phase boundary[29] acts as a red light modulator when subjected to UV optical gating (365 nm (3.4 eV)).

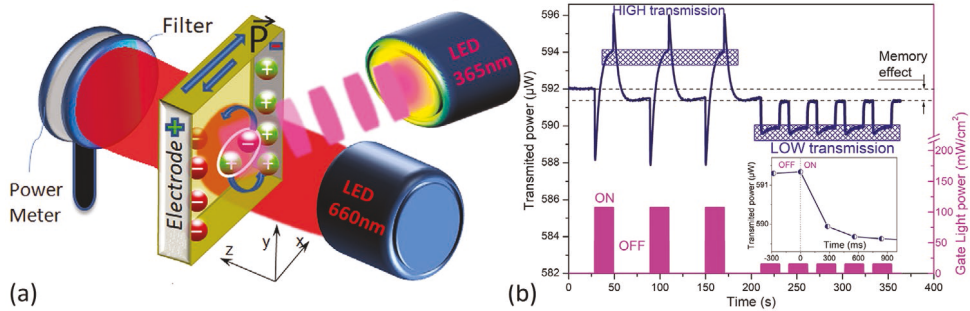

**Figure 1.** Optical control of light propagation. a) Schematic of the experimental setup with respect to the crystalline axis. b) Time dependence of the 660 nm beam transmission during periodic optical gating of 365 nm light. The inset shows a zoomed response of the depolarization regime limited by the LED response time (see text).

First the sample was electrically poled and its FE loop was recorded in darkness to ensure overcoming the ferroelectric coercive voltage. This action creates a homogeneous intrinsic electric field and forms a monodomain FE state known to improve light transmission[35,36] and PV properties.[37] The 660 nm light from a light emitting diode (LED) was then irradiated along the [001] direction and its transmission was measured as function of time while pulsing the UV light gate on the sample at 45° (Figure 1a). The propagation of the red light is successfully modulated with strongly irregular response involving both gain and loss in transmission signal with respect to the initial poled state. When the UV light gate is ON the transmission first decreases rapidly (Figure 1b (inset)) and then slowly increases following a saturation tendency. When switching off the gating light, the response reverses: the transmission increases rapidly and then relaxes more gradually to its equilibrium level. Moreover, for the first exposure to UV gating light a notable memory effect in transmission is observed (Figure 1b). For subsequent gating pulses the modulation of light transmission becomes reversible. Interestingly, the pulses with the same duration, but with different intensity, can trigger opposite transmission modulation. For the first three pulses (Figure 1b) of maximal intensity an increase in transmission is observed, while for the smaller intensities the transmission is globally reduced. This behavior occurs on a whole spectral range measured from 500 to 1000 nm (Figure S1, Supporting Information). The observed nonlinear response is also incompatible with dominant sample warming due to light absorption. Moreover, a light induced warming below 1.8 K was detected using Therm-app® infrared camera, which has a negligible effect on the light transmission (Figure S2, Supporting Information). In order to get more insight into the origin and dynamics of this complex photo-response, the sample was subjected to longer irradiation pulses and the electric current was monitored along with transmission and electric polarization (**Figure 2**). In the absence of gating light, the red illumination already causes a small background PV current of ∼-15 pA flowing along the measured direction. However, the situation changes drastically when gating light is ON. Large singularities occurring on different time scales are observed in the electric current measured along the previously poled [100] direction (Figure 2b). A large spike-like increase in the current over one order of magnitude is observed followed by a slower relaxation to the steady PV current level of ∼-65 pA (Figure 2b, inset). When gate light is off, this process is reversed. The transmission clearly correlates with the PV current. In the simplest picture, transmitted light can be scattered on free charges (elastic Thomson scattering) or electrons in the excited state can absorb light of a different energy than in the ground state (photochromic effect[38]). However, these effects should be linear with light intensity and independent on the direction of the current, which is not the case here. The observed charge dynamics clearly reveals the six specific regions marked as I, II, III, IV, V and VI on Figure 2.

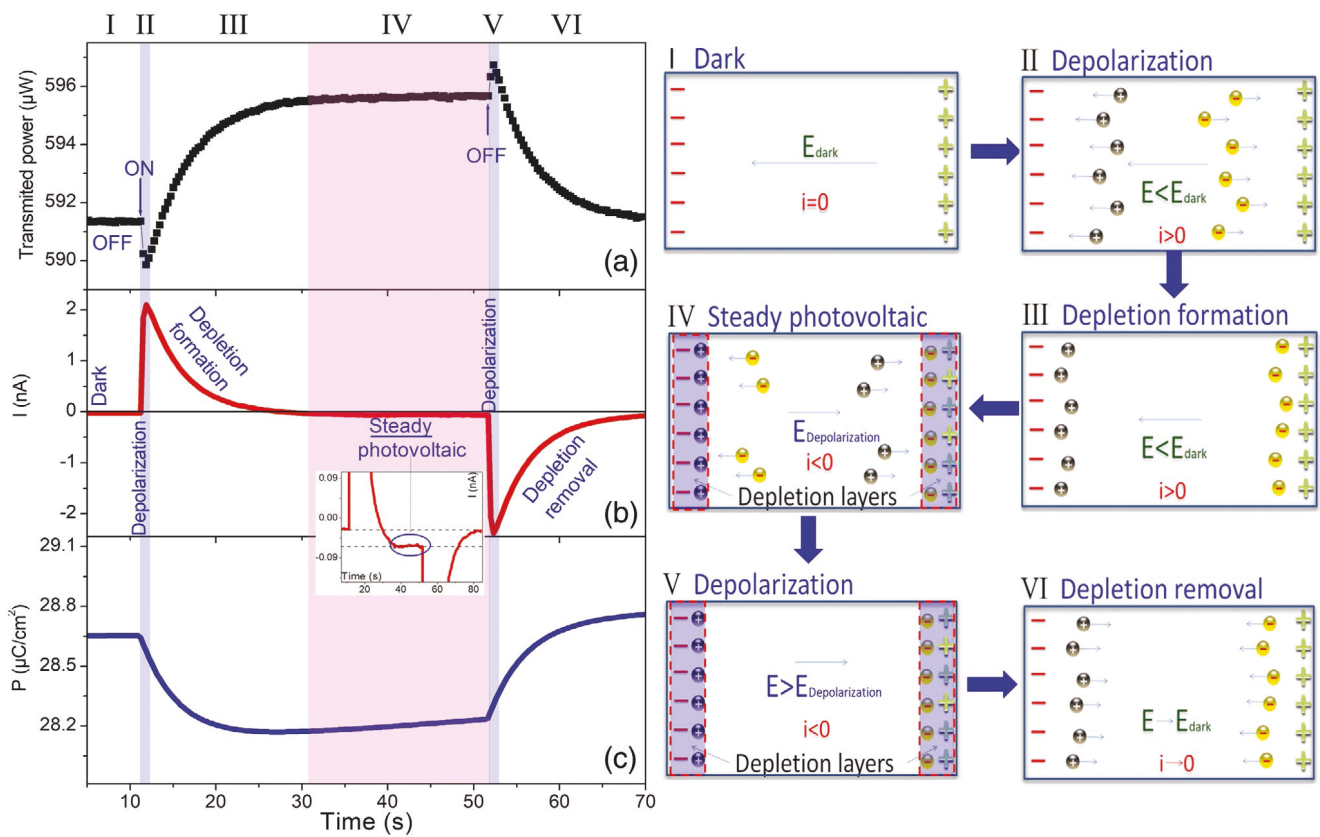

**Figure 2.** In-situ electro-optic measurements and modeling. a) transmission, b) photo-current and c) polarization. Regions I to VI are explained in the right part (see text).

In the ideal case in the previously poled FE there are no free charges in darkness (see "I" in Figure 2). When turning ON the UV light, free charges appear (see "II" in Figure 2) and reduce the stored electric field in agreement with the observed decrease of sample polarization (Figure 2c). The electric field attracts the charges to the edges forming the depletion layers.[39] The latter create a depolarization field of opposite direction (see model "III" in Figure 2). Therefore, the newly generated charges inverse their flow direction, in agreement with a sign change of the current in the steady PV region "IV". When light is off, there are no more charges to feed the PV current and the process reverses passing via depolarization "V" and depletion layers removal "VI" stages. It is also clear that the initial state may not be fully recovered explaining the observed remanent effect (Figure 1b). Therefore, the transmission modulation effect can be of electro-optic (Pockels effect[40]) origin where the UV gate light induces modifications in an internal electric field via the interplay between the depolarization and PV effects.

Indeed, the transmission response was confirmed to be electric-field-dependent and related to a butterfly-like hysteresis[41,42] with a clear voltage effect on the optical gating (Figures S3 and S4, Supporting Information). Both the irregular form of transmission response to regular square-like gate pulses and the intensity dependence of the effect (Figure 1b) strongly hint at large optical nonlinearity. Further insight can be gained by measuring the transmission as a function of gate light intensity (**Figure 3**).

The red light transmission presents a significant nonlinear response consisting of low transmission and high transmission regimes in agreement with Figure 1b. The transmission first decreases rapidly versus gating UV light intensity and then increases more slowly finally reaching a more than two times improvement with respect to the initial state. This effect correlates well with the photocurrent monitored across the [100] direction during the optical gating (Figure 2b). In fact, the two opposite contributions to the photocurrent are observed: the transient part (corresponding to depolarization) results in transmission decrease, and the steady photovoltaic component with relaxation has the opposite effect. The competition between these two mechanisms which occur at different time scales is likely responsible for the observed nonlinear response.

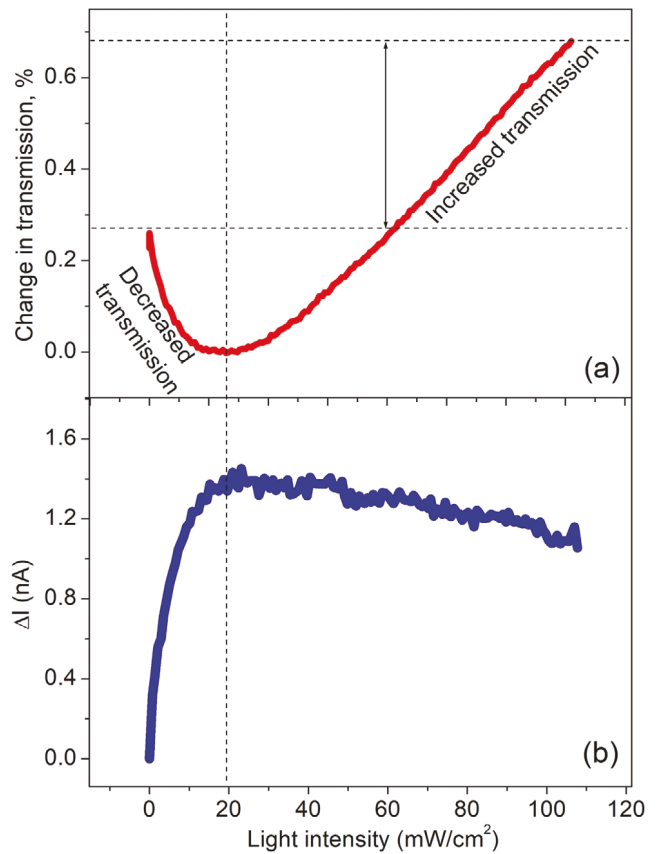

**Figure 3.** Intensity dependence. a) Red light (660 nm) transmission as a function of gate light intensity. b) In-situ measurements of electrical current along the [001] direction (Figure 1a). Light intensity sweeps, with a step of 2.36 mW cm$^{-2}$, taken at a rate of 200 ms.

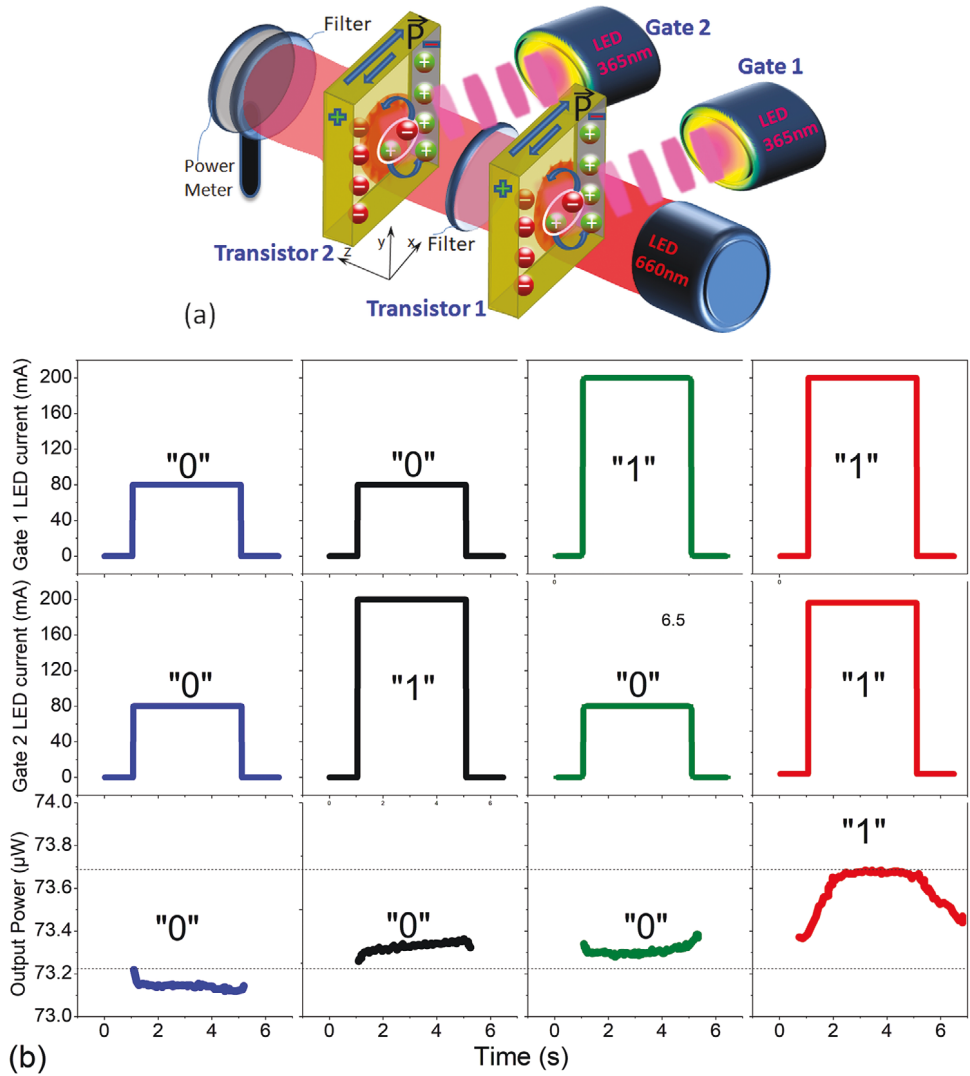

**Figure 4.** Demonstration of optical logic. a) Schematics of the experiment. The 660 nm light transmission is gated twice by 356 nm LEDs. b) Logic "AND" operation.

## 2.2. Optical Logic Operation

Because nonlinear dynamics is essential for general information processing,[43] we can take advantage of the observed pronounced nonlinearity to implement an optical logic operation as demonstrated in **Figure 4**. An additional optical transistor element was added in series to the transmission line, including its own optical gating LED element "gate 2" (Figure 4a). To illustrate a logic operation, the implementation of the Boolean logic function called "AND" is performed in Figure 4b. For this logic function the two binary inputs must be simultaneously ON (with state "1"), to result in the output being also ON (higher transmission state "1"). All other combinations of the input gate signals should result in a lower transmission (sate "0"). A complete logic operation consists of initialization, computation, and readout. The initialization involves stabilization of the transmission signal via the 2 crystals with previously adjusted intensities of each gate beams.

For demonstration purpose of the binary performance we selected the intensities of gate 1 and gate 2 to have a similar impact on the transmission signal, although multistate operation is also possible in agreement with Figure 1b. Next, the computing stage consists of inputting a combination of the two gate light pulses. Only when both gates 1 and 2 are ON, the increased transmission signal level is observed (state "1") which exceeds all other gating combinations (**Table 1**).

**Table 1.** Truth table for "AND" logic.

| Gate 1 | Gate 2 | Transmission OUTPUT |
| --- | --- | --- |
| 0 | 0 | 0 |
| 0 | 1 | 0 |
| 1 | 0 | 0 |
| 1 | 1 | 1 |

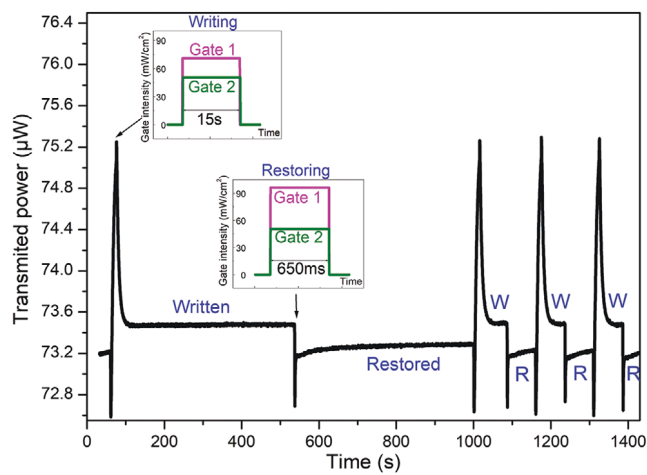

**Figure 5.** Optical logic level restoration. The long term 15 s exposure of the gates 1 and 2 induces a stable remanent effect that can be restored by shorter pulses of 650 ms duration. Insets show corresponding time profiles of the gating pulses.

### 2.3. Optical Logic Level Restoration

As the system is based on FE property, the remanent effects related to polarization state change due to charge trapping, spontaneous discharge or temperature drift may become a drawback. Luckily the FE state can be reset not only electrically, but also optically allowing us to achieve all-optical control of the device (**Figure 5**). This possibility can be realized optically if one takes advantage of the on-demand ability to increase or decrease the transmission (Figures 1b and 3). By adjusting the excitation energy and duration of the gating pulses, i.e., the amount of photo-carriers involved in the depolarization, we can independently increase or decrease the transmission of each element. Thus stable logical "1" states can be written with longer pulses (15 s), and reliably erased using short 650 ms pulses with optimized intensities, therefore demonstrating purely optical logic level control.

It should be noted that the response time is faster than the rise time of the LEDs used for illumination(∼450 ms), and that long time scales are shown to demonstrate stability. Using external modulation of the gating light, we observe an operating speed at least of 2 kHz (Figure S5, Supporting Information). Furthermore, by analogy to the FE polarization related photostrictive effect[44] the photoresponse should be faster in the materials with lower dielectric permittivity. Indeed, in the $BiFeO_3$ with much lower dielectric permittivity[45] the photoresponse can reach the THz regime.[46,47] In this regard other FE compounds with photopolarization properties[48] may become interesting candidates.

### 3. Conclusions

The potential impact and significance of the reported here new type of optical transistor effect can be assessed from the viewpoint of achieving light controlled operation of logic elements, meant to open the long awaited era of optical computing. In this regard, an optical transistor is a key element to make all-optical computing possible. Therefore, an alternative, straightforward and low-power device concept operating at ambient conditions is of primary importance. Moreover, the occurrence of a rewritable remanent transmission effect also demonstrates an attractive possibility to realize all-optical memory – a serious obstacle to the development of optical computing.[43] Our approach obviously satisfies many device maturity criteria including logic functionality and logic-level restoration. Thanks to the large transparency of the crystals, the signal light should be able to provide enough power to drive at least three sequential transistors. The presented device concept also provides input-output insulation, as the gate light can be easily filtered from the signal light since they are at different wavelengths. The reported optical gating effect can be understood through the competion between light-induced depolarization and a steady photovoltaic effect, which alter the transmission of light through the ferroelectric crystal. Our findings therefore open a new avenue to photonic control of optical devices based on photo-ferroelectric crystals for all-optical modulators, memories and variety of optical logics, possibly extendable even to quantum effects[49] and 2D structures.[50]

### 4. Experimental Section

The single crystals of $Pb[(Mg_{1/3}Nb_{2/3})_{0.70}Ti_{0.30}]O_3$ were from Crystal GmbH (Germany) and the exact composition of PMN-PT compounds was independently verified by energy-dispersive X-ray spectroscopy. The scan was performed under the electron beam accelerating voltage of 15 kV and take-off angle of 30°. The EDS spectrum confirmed the absence of significant amounts of impurities and the exact content was found to be in correspondence with the data of the supplier. The samples of 0.3 mm in thickness had (001) orientation with the edges' cut along [010] and [100] to form plates typically 1.8 mm × 3.5 mm. For FE poling both electrodes were formed with the silver paste covering the edges in the plane parallel to ZY plane. The PV properties (Figure S6, Supporting Information) were measured using a Keithley 6517B electrometer. Before transmission measurements the samples were electrically poled in darkness by sweeping the electric field from zero to −7 kV cm$^{−1}$ followed by −7 to +7 kV cm$^{−1}$ sweep and then back to zero. The signal illumination for transmission experiments was generated by a light emitting diode of 660 nm with 18 nm bandwidth. The optical gating were performed by fibered LEDs of 365 nm selected to be above to the bandgap of the crystal (3.23 eV).[51] These LEDs had a rise time of 450 ms in a constant current mode and bandwidths of 9 nm (max. power 126 mW) and 7.5 nm (max. power 54.7 mW) for gate 1 and gate 2 respectively. The signal beam was filtered after each gate by the corresponding red light optical filters. The 520–1100 nm transmission spectra were measured using a Halogen lamp from an AvaLight-DH-S-BAL source, whose emission spectrum lies above 500 nm, and an AvaSpec-ULS2048XL-RS-EVO-US spectrometer with a 300 lines mm$^{−1}$ grating and 25 μm sized slit. A 4 mm diameter sample area was illuminated using a 400 μm diameter broadband fibre coupled to a confocal lens. The transmitted light was collected using a second lens and fibre. Drift correction was performed before and after using bifurcated fibres and a multiplexer to switch between the sample path and a direct light source-to-detector path. Dark and reference spectra were acquired for 1.8 ms integration time with 1000 averages between 520 to 1100 nm, while time-dependent spectra were acquired every 42 ms and were averaged 10 times. The sample was irradiated with UV light using a Thorlabs M365FP1 LED at a power density of 110 mW cm$^{−2}$ for 30 s.

## Supporting Information

Supporting Information is available at the end of this preprint.

## Acknowledgements

Partial support from the Labex NIE Grant No. 0058_NIE and Ph.D co-fund program of A.M. of the Alsace region is acknowledged. M.V. knowledges EU Horizon 2020 research and innovation program under grant agreement N° 766726, project TSAR and the 'projet transversal ELSA' from CEA as well as French National Research Agency (ANR) through the SANTA project (18-CE24-0018-01).

## Conflict of Interest

The authors declare no conflict of interest.

## Author Contributions

B.K. has initiated the experiment of the optical gating of light transmission. The measurements were carried out by A.M. under supervision and participation of B.K. Samples were analyzed by R.G. and U.B. using energy-dispersive X-ray spectroscopy. P.D. and B.K. performed transmission spectra measurements. M.V., B.D., J.-F.D. and R.G. participated in discussions and manuscript writing.

## Keywords

ferroelectrics, light transmission, optical materials, optical memory, optical transistor

Supporting Information

Photovoltaic-Ferroelectric Materials for the Realization of All-Optical Devices

*Anatolii Makhort, Roman Gumeniuk, Jean-François Dayen, Peter Dunne, Ulrich Burkhardt, Michel Viret, Bernard Doudin, and Bohdan Kundys\**

# Supporting Information

**Photovoltaic - ferroelectric materials for the realization of all-optical devices**

*Anatolii Makhort, Roman Gumeniuk, Jean-François Dayen, Ulrich Burkhardt, Peter Dunne, Michel Viret*, *Bernard Doudin* and *Bohdan Kundys\**

Figure S1 illustrates transmission spectra of the $Pb[(Mg_{1/3}Nb_{2/3})_{0.70}Ti_{0.30}]O_3$ crystal for different gate light intensities obtained using a white light source halogen lamp illuminated along the [001] direction. The non-linear response involving both transmission decrease and increase depending on the gate light intensity is observed.

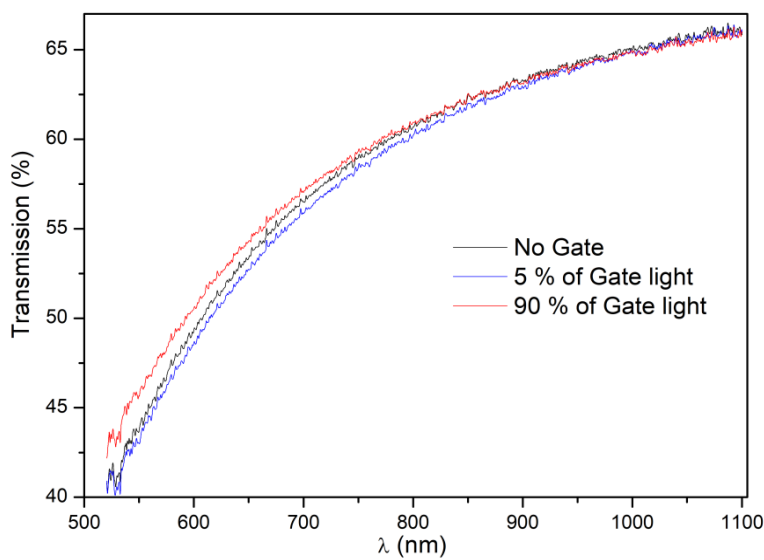

Figure S1. Spectral response of the photo-gated light transmission in spontaneous FE state.

Figure S2 compares the red light transmission response to UV gating light and to the sample warming by electrical heater. The influence of UV gating light illumination on the sample temperature was determined using infrared camera to be less than 1.8K. When sample is warmed by this amount using the electrical heater the change in transmission is negligible. Even if the sample is warmed by ~7K, only a small increase in transmission is observed. Moreover, the warming does not decrease the transmission; the change is linear and much smaller as compared to the UV light induced effect.

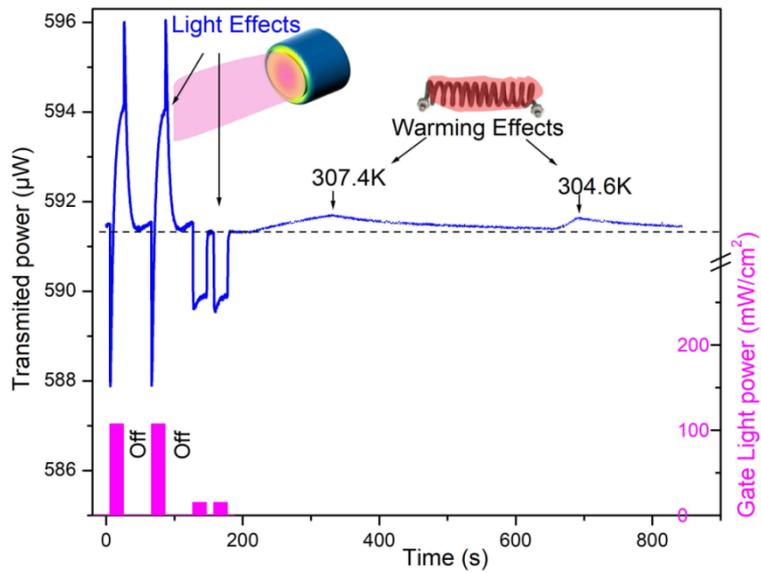

Figure S2. Comparison between light induced and thermal effects.

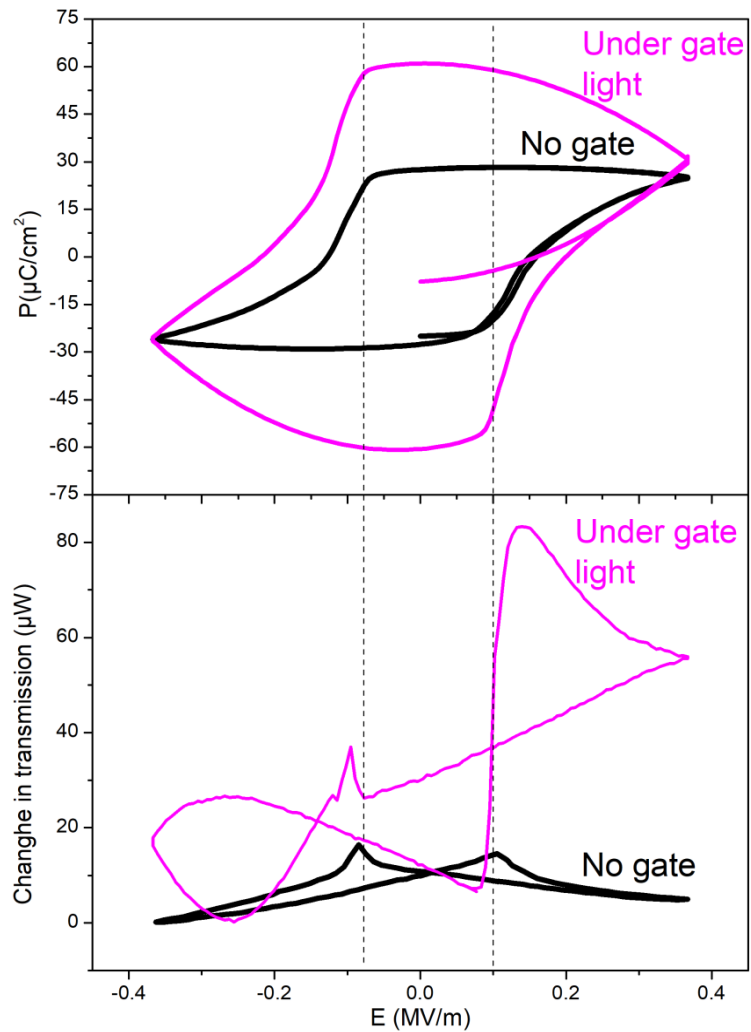

Figure S3. Comparison between FE (upper panel) and transmission loops (lower panel) under UV gating and in darkness.
2

Figure S3 shows ferroelectric loops in darkness and under UV gating (upper panel) with transmission measurements (lower panel). In the absence of gate light the transmission shows a „butterfly-like" hysteresis loop clearly correlating with the FE hysteresis. Such a behavior is usually seen in other FE materials[42] and is linear for subcoercive voltages constituting a Pockels effect. However, when the UV gating is ON this behaviour drastically changes because free charges modify the electric field inside the sample. Indeed, the new shape of the transmission loop induced by the gate UV light resembles the one previously observed for the electrode induced electric field modification[51]. Importantly, the transmission in this case has two remanent states implying an interesting photoferroelectric memory storage possibility.

Figure S4 additionally illustrates the influence of a static electric field on the time dependent transmission, steeming from a linear character of the Pockels effect for subcoercive ferroelectric region.

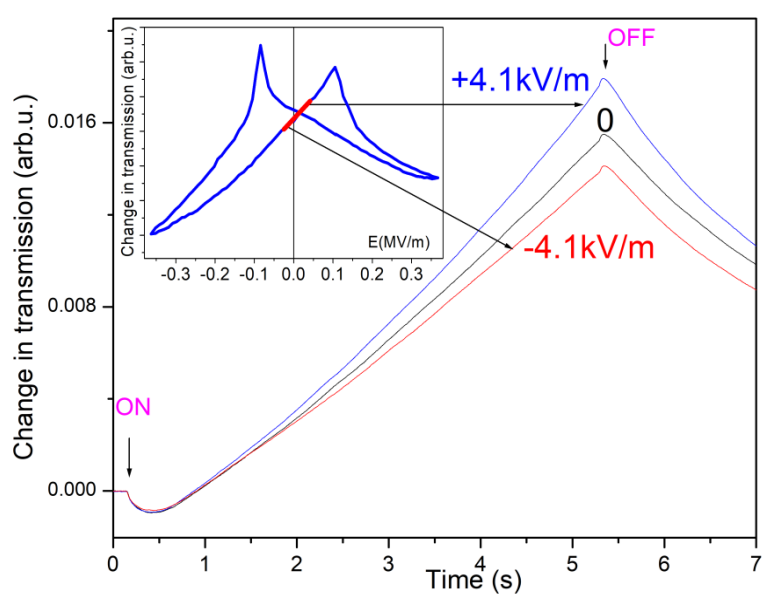

Figure S4. Impact of the applied electric field on the UV gated light transmittance. Inset shows the corresponding subcoercive electric field region.

To test the response time speed, the transmission was monitored as a function of time while modulating the gating light at different frequencies. The response is found to follow the modulation frequency at least up to 2 kHz and may have a room for farther improvement with the optimization of the experimental setup, sample material and geometry.

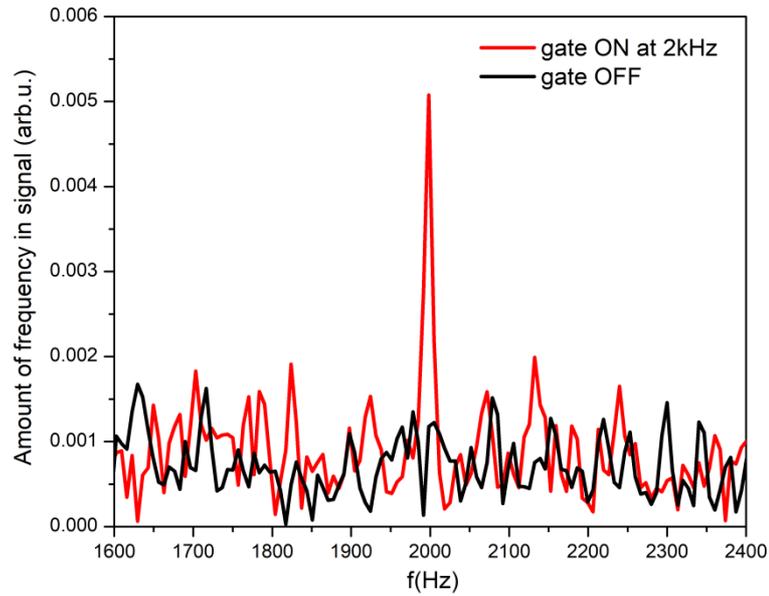

Figure S5. Fourier transform of the transmitted red light signal in response to a sinusoidal UV gate light modulation of 2 kHz.

Figure S6 depicts the current-voltage measurements of the PV properties of the sample in the same geometry as shown in fig 1b. The I-V curves were recorded for sequential light intensities increase without overcoming coercive voltages. Both short-circuit current ($I_{sc}$) and open circuit voltage ($V_{oc}$) change with light intensity and follow saturation tendency.

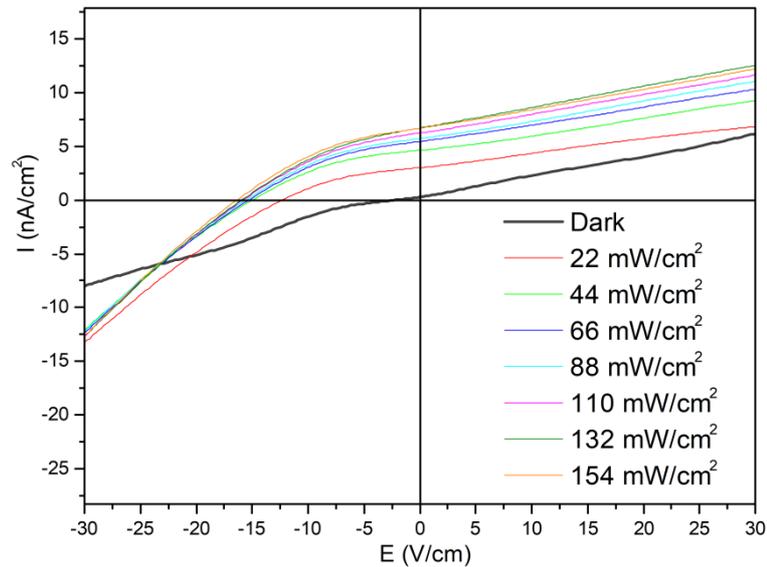

Figure S6. Current-voltage measurements of the PV properties of the Pb[(Mg$_{1/3}$Nb$_{2/3}$)$_{0.70}$Ti$_{0.30}$]O$_3$ crystal excited with 365nm LED light.